\documentclass[12pt]{article}

\def\beq{\begin{equation}}
\def\eeq#1{\label{#1}\end{equation}}
\def\eeqn{\end{equation}}

\def\beqa{\begin{eqnarray}}
\def\eeqa#1{\label{#1}\end{eqnarray}}
\def\eeqan{\end{eqnarray}}

\let\bar=\overbar

\def\Dslash{\not{\hbox{\kern-4pt $D$}}}
\def\dslash{\not{\hbox{\kern-2pt $\del$}}}

\def\msb{{\bar{\ssstyle M \kern -1pt S}}}

\usepackage{fancyhdr,graphicx}
\def\Title#1{\begin{center} {\Large {\bf #1} } \end{center}}

\begin{document}

\Title{The Figure of the Sun from Ground-based Experiments}

\bigskip\bigskip


\begin{raggedright}
{Costantino Sigismondi\index{Sigismondi, C.}\\
\it International Center for Relativistic Astrophysics,\\
P.le Aldo Moro 5 00185, Roma, Italy, \& \\
Observat\'orio Nacional, \\
Rio de Janeiro, Brazil.\\
{\tt e-mail: sigismondi@icra.it}}

{S\'ergio Calderari Boscardin\index{Boscardin, S. C.}\\
\it Observat\'orio Nacional,\\
Rua General J. Cristino 77 20921-400 Rio de Janeiro, Brazil.\\
{\tt e-mail: sergio.boscardin@on.br}}
\bigskip\bigskip
\end{raggedright}

\bigskip\bigskip

\abstract{The figure of the Sun reflects its inner structure and dynamics, influencing also the perihelion precession of close orbiting bodies, like Mercury. 
To study the solar figure from ground, the deformation on the solar image induced by the atmosphere has to be known up to one part over a million, and this is done through differential refraction models: a historical review of them is drafted.
The solar oblateness has been investigated in order to validate alternative theories to General Relativity and to understand the internal dynamics of our star.
The solar figure should possess only micro departures from sphericity according to the standard stellar structure theory and helioseismology data,
though variations along the cycle has been observed. 
Ground-based and satellite data show contrasting observational results. 
The oblateness measured onboard RHESSI satellite, the one of SDS onboard a stratospheric ballon and that one of the Astrolabe of Rio de Janeiro are presented 
with their implications in classical and relativistic gravitation.
The perspectives offered by the reflecting heliometer in the future measurement of the oblateness are depicted.}

\section{Introduction: the atmospheric refraction and its comprehension along the history}

\begin{figure}[htb]
\begin{center}
\includegraphics[width=15cm]{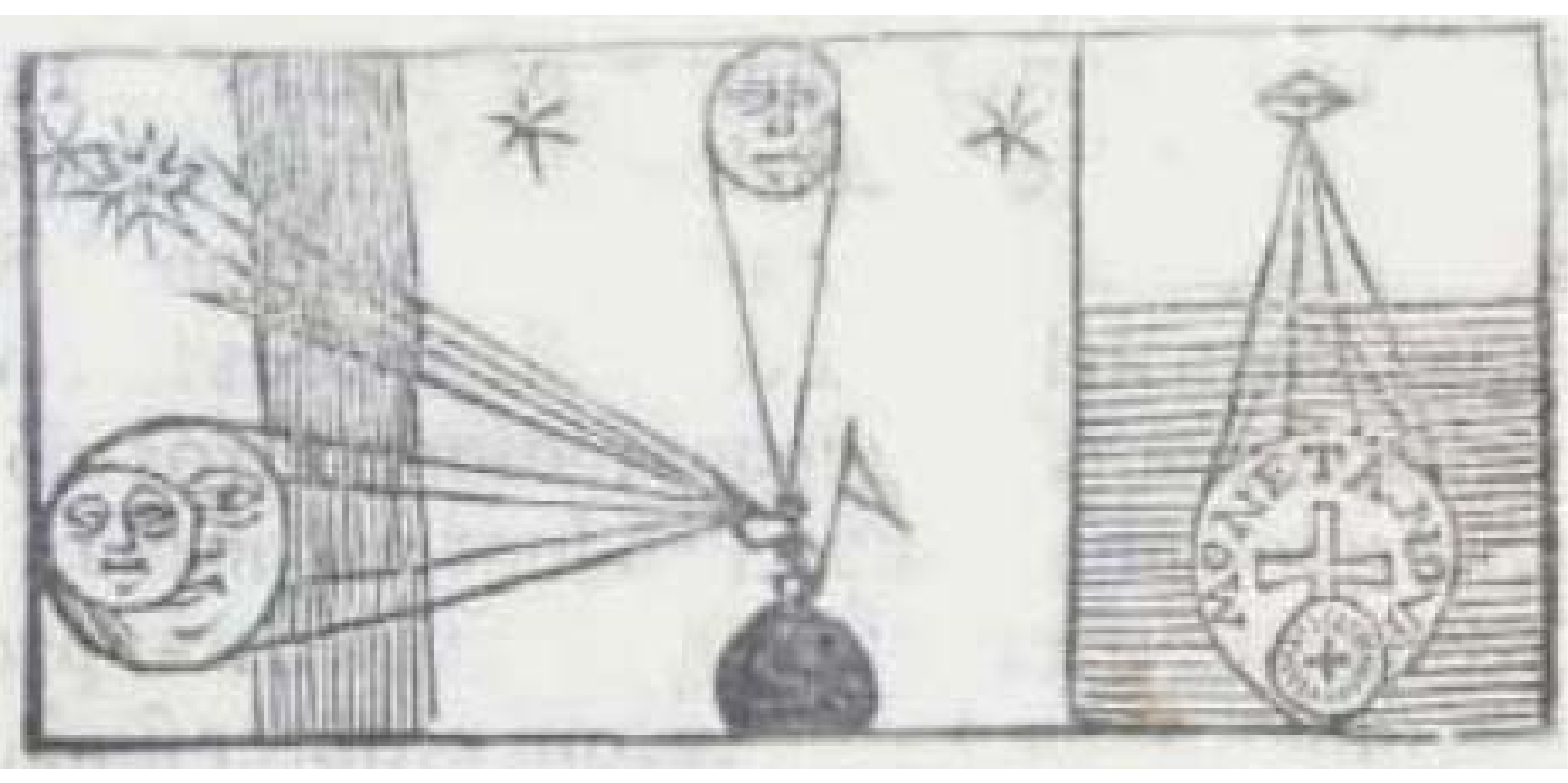}
\caption{The observed dimension of celestial objects near horizon and at the meridian, as in book I chapter VI of the Sphere of John Holywood.}
\label{fig:uno}
\end{center}
\end{figure}

The influence of the atmospheric refraction on the image of the Sun 
has been argued since antiquity, with Ptolemy (about year 150), Ibn Sahl\cite{sahl} (984) and Alhazen (1020). The European reception of these optical phenomena in the first Renaissance occurred thanks to the work of the polish scientist Witelo who wrote the {\sl Perspectiva} (1278).
The {\sl Sphaera} of John Holywood (1256) is a manual of introductory astronomy used in Paris University with a large editorial success. Famous is the {\sl Commentarius in Sphaeram} made by Clavius (1570, and several editions in the following 30 years).
In the {\sl Sphaera} the problem of the appearance of solar figure near the horizon or at zenith is presented in this way: near the horizon the Sun and the Moon are larger than at zenith, see Fig~\ref{fig:uno}.

The explanation is by analogy: as a coin seen through pure water appears bigger in the same way the Sun and the Moon are bigger because their luminous rays pass through thick vapours.
Therefore in the {\sl Sphaera} the refraction is introduced, but only qualitatively.
Clavius\cite{clavius} did not retain this explanation, about the variation of the angular dimensions, but only the discussion about the appearance of the Sun and the Moon near the horizon before they are actually there. This explanation is correct, but the differential effect, stronger for the lower limb, is not yet mentioned.  
The quotation of the coin seen into the water, as reported by Clavius, is from Alfraganus (Al-Farghani, {\sl Elements of astronomy on the celestial motions} about 833), the magnification effect was probably due to the curvature of the bottle used for the experience. Actually in the case of the atmosphere we do not have such curvature, nor the magnification at horizon, which happens to be a psychological effect due to the proximity of the Sun or Moon to earthly objects, like trees, houses, mountains... which are considered big, while in the middle of the sky there is no comparison with usual objects.

The correct formulation of the atmospheric refraction was achieved after Tycho's observations of the
supernova in Cassiopeia (1572) and with the law of Snell (1619).
The controversy about the authorship of the law of refraction includes also Abu Said Al-Ala Ibn Sahl, Thomas Harriot, Willebord Snell, Ren\'e Descartes.\cite{sahl}
The application to the atmosphere, as a refractive medium, is known as the Laplace law, referring to his {\sl M\'echanique C\'eleste} (1799-1805).
In the theory of atmospheric refraction Laplace shows that the ratio $sin (i)/sin (r)$ of incidence and refracted angles is constant as in the Snell law.\cite{belhoste}
The parameters used to implement the Snell-Descartes law were confirmed by the experimental measurements made by Gay-Lussac onboard a balloon flown up to 7000 meters in 1804 ({\rm http://www.chemheritage.org/discover/online-resources/chemistry-in-history/themes/early-chemistry-and-gases/gay-lussac.aspx}).
The Laplace's\cite{laplace} historical work of 1825, the fifth volume of the {\sl M\'echanique C\'eleste}, resumes all contribution to this science in the last century, for this reason the law of atmospheric refraction bears his name, but a complete treatment of atmospheric refraction was already available in the {\sl Trait\'e Des mouvemens apparens des corps c\'elestes}(1786) of Dionis du S\'ejour, with the refraction index evaulated in $r=1.00028=3840/3839$.
In particular the vertical and the horizontal components are there explained in detail.

\section{The horizontal atmospheric refraction and the apparent solar diameter contraction}

A first experiment on the measurement of the solar diameter was realized by Archimedes: it was made at sunrise and sunset, with an horizontal rod and a moving cylinder.\cite{sigi05} 
The experiment has not been replicated at other solar elevations and along different heliolatitudes. Only the horizontal diameter was measured, upper and lower limit, and the value was 1/720 of $360^o$, say 30 arcminutes.
As we have seen in XVIII century the mathematical formalism to treat differential refraction was already complete, and 
the horizontal component of the refraction was quoted by Danjon (1950, p. 156); the contraction of the solar diameter for the horizontal differential refraction is about 0.6 arcsec.\cite{danjon}

To understand the horizontal refraction we exploit the example of Rio de Janeiro, where the Sun's center transits onto the zenith on 9th December and 1st January, and for more than a month the Sun culminates very close to the zenith, with the minimum differential refraction vertical component. 

On the zenith this component, for symmetry reasons, corresponds also to the horizontal differential refraction, 
that is the same for all altitudes, as reported in the text of Du S\'ejour\cite{dusejour} (1786, p. 243): it contracts the solar radius of 0.25 arcsec.

A simple derivation of the value of the horizontal differential refraction is made with the Snell law around zenith for the solar radius,
by using the refractive index of the atmosphere n=1.00027784 at 550 nm, sea level and STP conditions: standard pressure and temperature.

\begin{figure}[htb]
\begin{center}
\includegraphics[width=15cm]{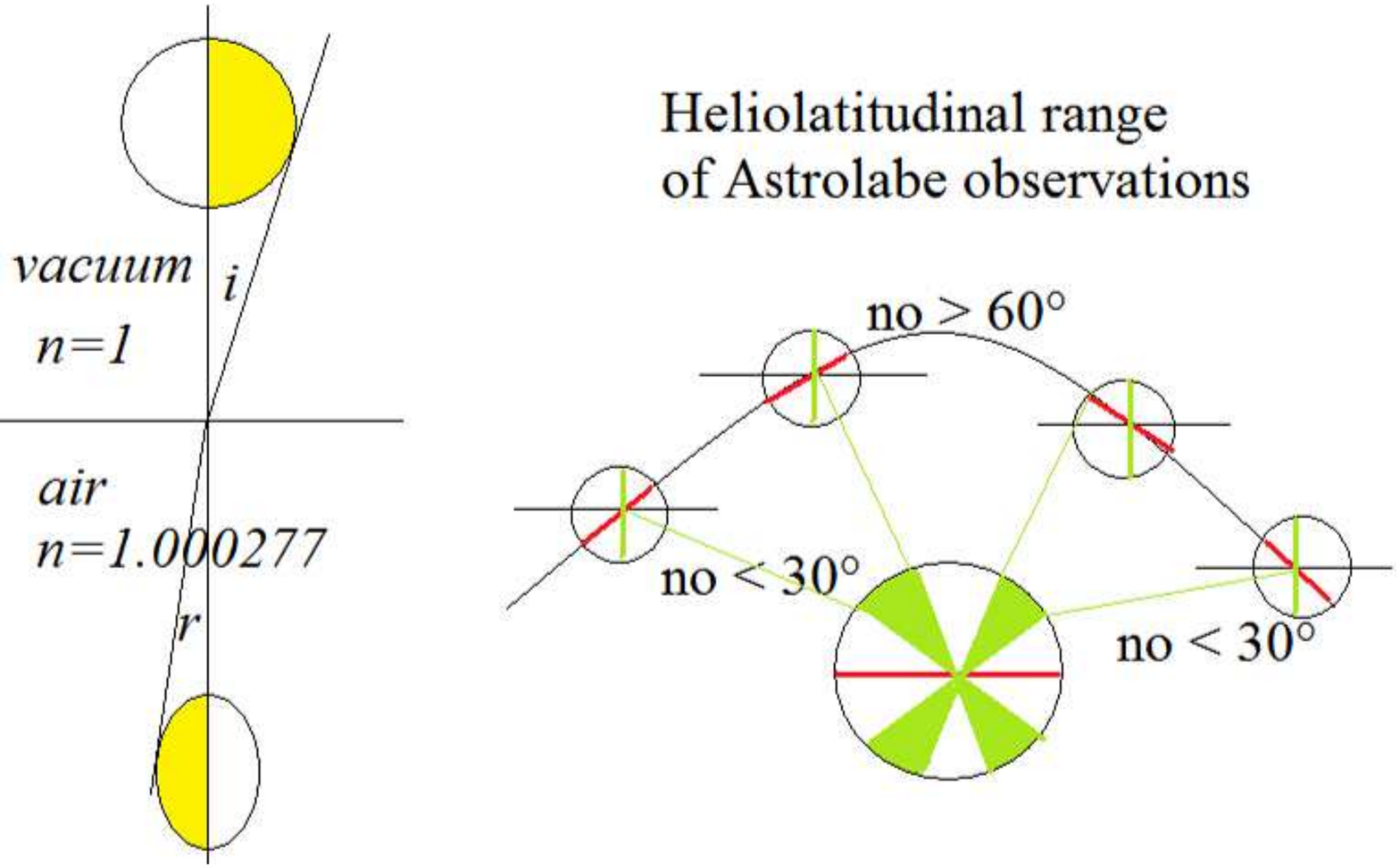}
\caption{The geometry of atmospheric refraction in the case of Sun at zenit. On the right side the heliolatitudes observed during an observational session with the astrolabe, limited to $z=30^o \div 60^o$. The number of observations per interval of heliolatitude confirms this deduction and it is shown in the Fig~\ref{fig:sigma}.}
\label{fig:snell}
\end{center}
\end{figure}

 {\rm http://refractiveindex.info/?group=GASES\&material=Air}
The Observatory of Pulkovo published the Refraction tables in 1870, 1905, 1930, 1956 and 1985.\cite{abalakin} 
A Ray tracing in the modified US1976 atmosphere has been developed to take into account anomalous refractions in extreme conditions, like polar sunsets.\cite{werfe}
We consider the solar diameter of 32 arcminutes, or 1920 arcsec. We pose the Sun center in the zenith, and we calculate the position of the observed limb after the refraction. Applying the Snell law to the solar radius:
$sin(i)=1.000277\times sin(r)$ (see Fig~\ref{fig:snell} left-side)
$sin(960")=1.000277\times sin(r)$ being small angles $sin(r)\sim r = 960/1.00027784" = 959.733"$
the solar radius is reduced of 0.267" and the solar diameter of 0.533" in agreement with Danjon and Du S\'ejour values. 

While the solar astrolabes measured the vertical solar diameters since 1975, the possibility to verify by observations the horizontal refraction effect was possible with meridian and hourly circle transits, measured since the XVII century with increasing accuracy. 
The use of videorecording devices since 1990 allowed to reach subarcsecond accuracies, needed to see the horizontal refraction, in transits mesurements.

All types of heliometric measurements, where the whole figure of the Sun is studied, are subjected both to the effects of vertical and horizontal refraction.

\section{Vertical differential refraction for heliometric measurements}

\begin{figure}[htb]
\begin{center}
\includegraphics[width=15cm]{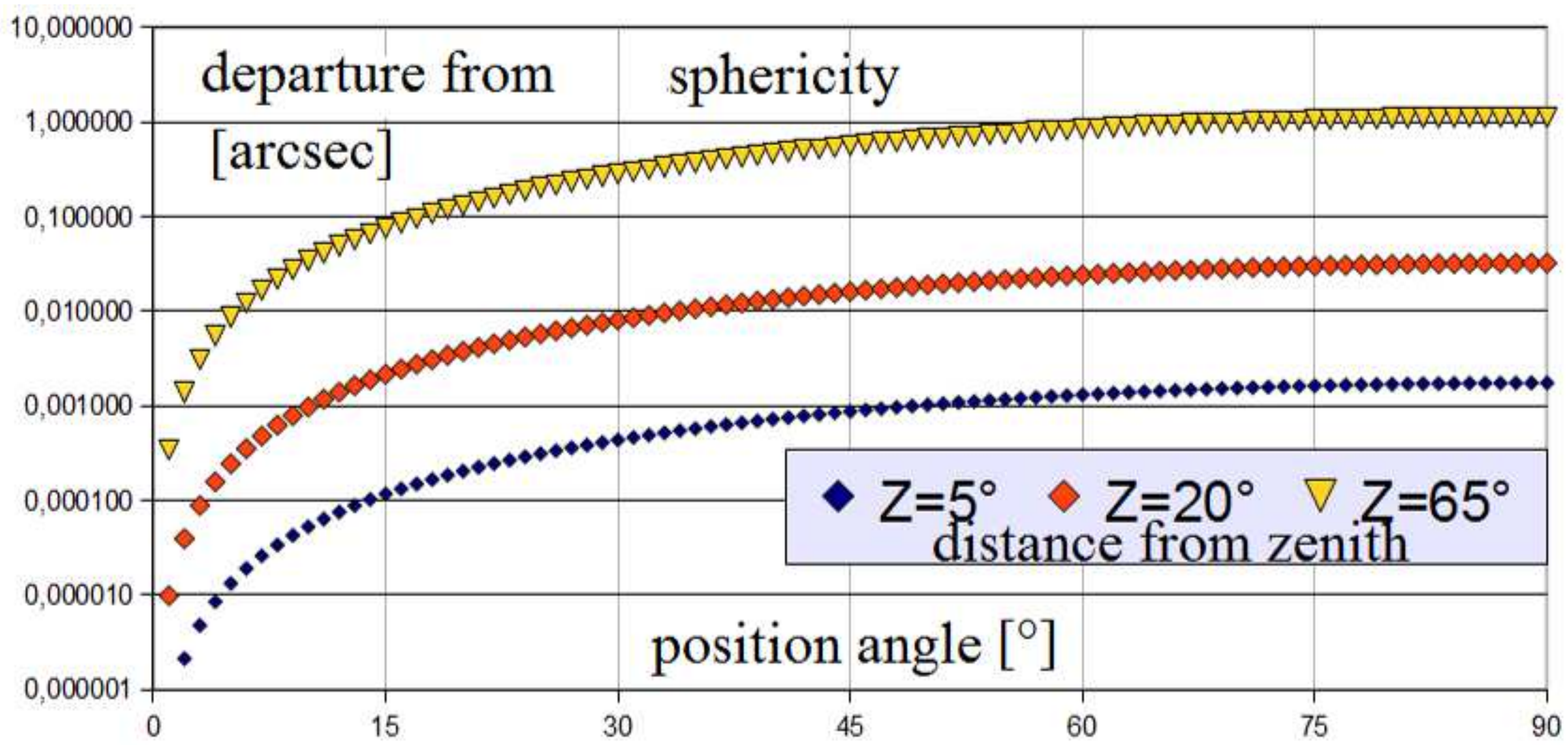}
\caption{The departure from sphericity in the case of  $25^o$;  $70^o$ and  $85^o$ of altitude, or zenithal angles $z=65^o; 20^o; 5^o$.
The model of atmospheric refraction of Penndorf\cite{penndorf} is included in the calculation. 
The departure of sphericity plotted here is a sort of flatness $f$: the ratio of the difference between the horizontal radius of the Sun and the one at a given angle $\theta$ with respect to the horizontal radius $a=(R_h-R_{\theta})/R_h$.}
\label{fig:excel}
\end{center}
\end{figure}

Implementing in a worksheet the formulae of Laplace (1805) with modern upgrades\cite{werfe,penndorf}for the vertical refraction and the horizontal component
we calculate the departure from sphericity of the solar image.
Conversely it is possible to use these values to deconvolve the result of heliometric measurements.
In Fig~\ref{fig:excel} there are the departures from a circle as function of the Position Angle PA, measured from W/E limb ($PA=0^o$) to $PA=90^o$ corresponding to the  upper/lower limb, in order to reproduce the case of a meridian transit at $70^o$ meridian transit (summer solstice in Rome), or the case of a transit at $25^o$ (winter solstice in Rome), and for $85^o$ (culmination from November to February in Rio). The Sun in Rio is observed at all these altitudes with the Reflecting Heliometer. It is to be noted that the definitions of asphericity, oblateness and flatness are slightly different and have not to be confused.\cite{rozelot}

\section{The influence of the turbulence on the solar image}

The atmospheric turbulence acts on the observed position of the inflexion point.
It is quantified in 0.08 arcsec for each arcsecond of Fried parameter,$\rho_0$\cite{sigitesi} or 0.08 arcsec for $\rho_0=$1.5 arcsec.\cite{desnoux}  
The astrolabe of Rio de Janeiro worked averagely at $R_0\sim40~$mm\cite{boscardin11} which in arcsec corresponds to $\rho_0\sim$2.5 arcsec; 
therefore the inflexion points of the solar images observed in Rio were, in average, shifted toward the center of 0.09 arcsec. 
Consequently the observed solar diameter was 0.18 arcsec smaller than the real Sun, because of the turbulence.
Each measured diameter has been corrected by that turbulence effect. 

Anomalous refractions are also invoked to explain different shifts of zenithal stars observed along the night.\cite{Taylor}
According to our measurements with the Sun, we can classify such anomalies among the low frequency motions of the atmosphere, with frequencies below 0.01 Hz and amplitudes of several arcminutes. The whole image of the Sun is moved by such motions, determining arcseconds-size scatters among meridian transit measurements of the solar diameter.
 

\section{Solar figure, General Relativity and interior solar dynamics}

The solar oblateness was investigated carefully in the last two centuries: after verifying the absence of an {\sl intramercurial} planet only a quadrupole moment of the Sun could explain the precession of Mercury's perihelion within the Newtonian gravitation.
After the onset of General Relativity (GR) in 1916 which explained the precession without the oblateness, in the 1960s this parameter was measured with great care to define alternative gravitational theories to GR by Robert H. Dicke and collaborators in Princeton.\cite{dicke}Dicke realized experiments to measure the oblateness of the Sun, based upon high frequency measurements of luminosity excesses at 
various position angles, modulated by a chopper rotating at opportune angular speed. Later Hill and Stebbins developped SCLERA to measure the oblateness.\cite{sclera}
Accurate discussions on the effects of solar spots (depression of the limb, also called Wilson effect, as in the case of the spot at 1 arcmin from the solar limb on 23 Aug 2002 h 14 UT) and faculae\cite{ingersoll} to the solar figure appeared in the literature after these pioneering experiments, as the studies on asymmetry and variations of solar limb darkening along the diameter measured by diurnal motion (drift-scan)1.\cite{neckel}   
Dicke and collaborators achieved an accuracy of $10^{-5}$ on oblateness measurements.
J.P. Rozelot and collaborators continued the studies at Pic du Midi Observatory 
aiming to uncover the internal structure of the Sun.\cite{rozelot}
They used the scanning heliometer, set up there in 1970-1996 by Jean R\"osch.
Rozelot, Lefebvre and Desnoux\cite{desnoux} during a series of days with excellent seeing achieved an accuracy of $10^{-6}$ 
on their oblateness measurements and determined quadrupolar and hexadecapolar gravitational moments of the Sun. After that experience,
the authors retain that only excellent seeing is worth for performing such measurements and, moreover, all measurements of solar oblateness made before 1996 have a mere historical importance.

\begin{figure}[htb]
\begin{center}
\includegraphics[width=15cm]{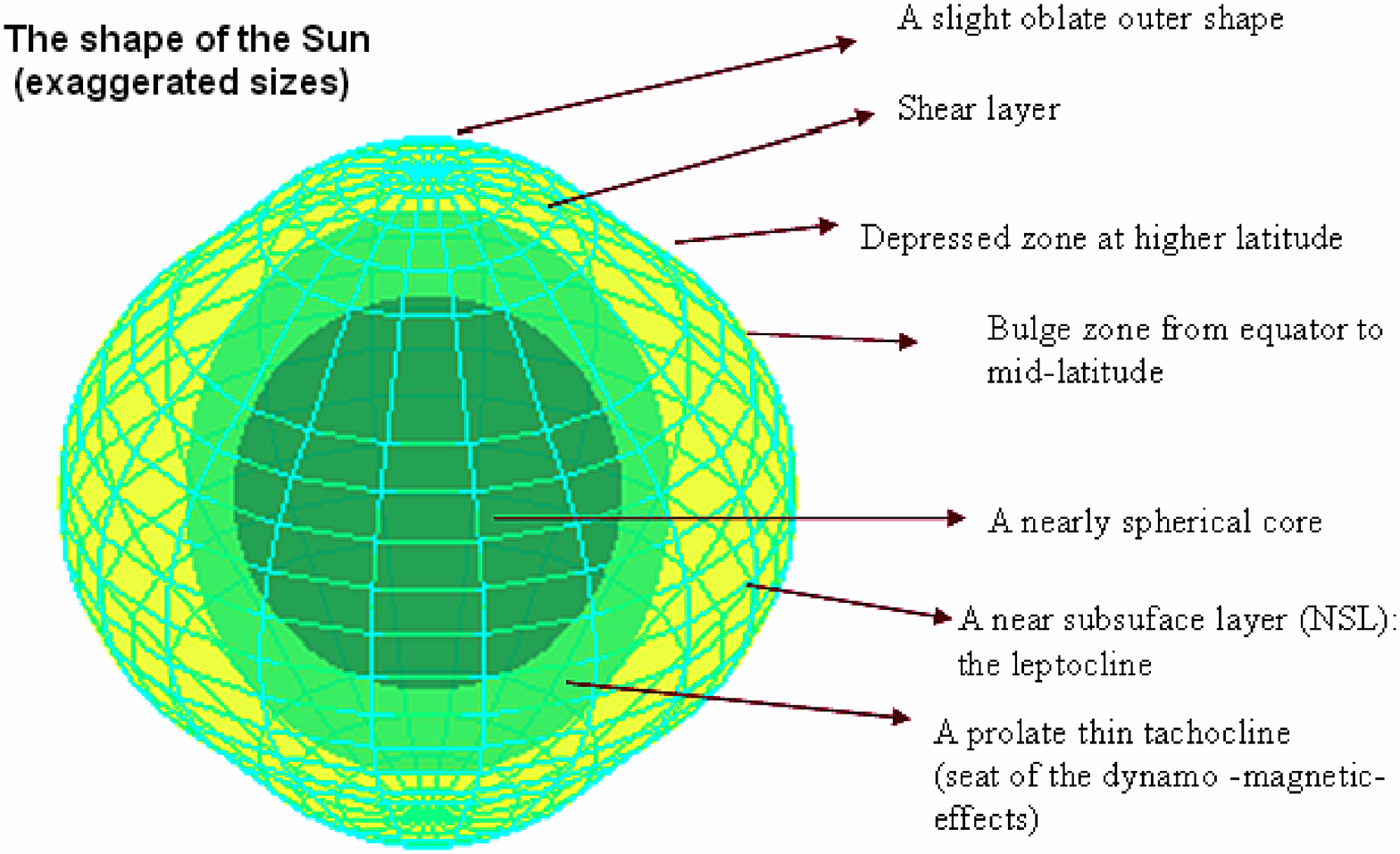}
\caption{The oblateness and other departures from sphericity, from Rozelot et al. (2003):the Sun includes a spherical solar core rotating at a nearly uniform velocity rate, a prolate solar tachocline and an oblate surface shape, both rotating at different velocities; it results
mainly from latitudinal shears and thermal winds on the surface, that affect the surface which is thus
corrugated.}
\label{fig:rozelot}
\end{center}
\end{figure}

The Fig~\ref{fig:rozelot} shows a sketch of the deformations of the solar figure.

Contemporarily the balloon-borne Solar Disk Sextant in several flights at 37 Km of height under 3 mb of Earth's atmosphere,
performed some measurements of the solar oblateness again in the range of some parts in $10^{-6}$.\cite{sofia2013}

A small instrument of 4 cm onboard RHESSI satellite recently evidenced an equator-to-pole radius difference of 10.77 mas,\cite{rhessi}
with a polar flattening of 6 Km over 700000 Km of solar radius.
It was interpreted as a consequence of the slow solar rotation 8.01 mas, and to magnetic elements in the enhanced magnetic field. 
This measurement as accurate as 1 milli-arcsec has been obtained 
well beyond the limit of diffraction of 2.5 arcsec for such instrument.
RHESSI was'nt designed to measure the oblateness and this result was serendipitous. 
This can be criticized, in the same way the results of SOHO satellite about the solar diameter variations have been criticized.
\cite{delmas}

\section{Oblateness with Rio de Janeiro astrolabe}

\begin{figure}[htb]
\begin{center}
\includegraphics[width=15cm]{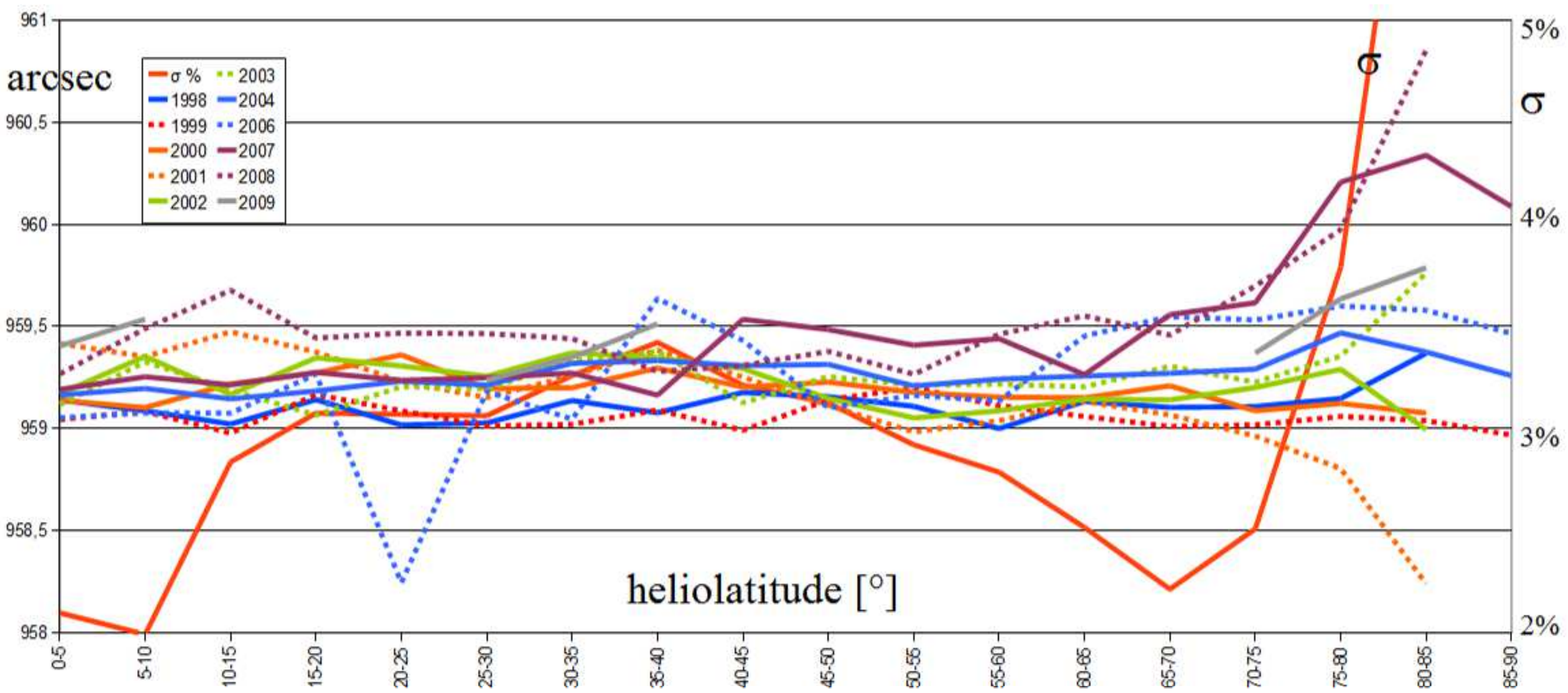}
\caption{The oblateness at all heliolatitudes during each year of observations. There are great differences from an year to another. 
These may not be solar features, since an oblateness of $\epsilon  =10^{-5}$ 
corresponds to 0.8 arcsec/cy of perihelion precession of Mercury, and actually only 0.15 arcsec/cy remain unexplained.
The statistics on the number N of observations available as function of the heliolatitude is presented through the parameter 
$\sigma\%=1/\sqrt{N}$.}
\label{fig:sigma}
\end{center}
\end{figure}

The solar astrolabes invented in the XX century, exploited accurate timing measurements to study the vertical diameter of the Sun.
The observations made with the solar astrolabe are characterized in the following ways:

\begin{itemize}
\item{the high frequency atmospheric turbulence contracts the diameter; it is measured during 20 s around the contact instants of the images; during these 20 s the seeing is also measured, to take into account its effect on the measure}
\item{the low frequency atmospheric motions move the whole image of the Sun and modify the contact times and the measured diameter}
\item{the vertical refraction does not act on the image since it drifts in the field of view of the telescope, and the vertical dimension is measured through the timing
at the same altitude (same almucantarat circle). As in the case of the solar diameter measured at sunset\cite{sigi-sunset}, near the horizon the Sun slows down its descent, the angular velocity is slower in the preceding limb than in the proceding, but the total timing is the same as in absence of the atmosphere, because when the proceding limb touches the horizon it has been slowed down by the atmosphere in the same way as occurred for the preceding at its contact.}
\item{the horizontal refraction is the same at each altitude,\cite{dusejour} it acts on the timing across meridian and hourly circles 
because it reduces the lenght of the horizontal chords, but for vertical diameters it does not act.}
\item{the latitude of the observer determines the range of heliolatitude studied by the astrolabe: the Sun describes a circle in the sky during the day, and the angle described with the horizon is approximately equal to the co-latitude $\theta=90^o-\lambda$. 
The intersection with an almucantarat occurs at an angle ranging between $\pm\theta$; considering that when the Sun is lower than $30^o$ above the horizon and higher than $60^o$ there are no observations with the astrolabe, the range of heliolatitudes observable from a given location is limited to $\sim\pm\theta/2$, in favour of tropical locations, like Rio de Janeiro (see Fig~\ref{fig:snell} right-side and Fig~\ref{fig:sigma}).}

\end{itemize}

\begin{figure}[htb]
\begin{center}
\includegraphics[width=15cm]{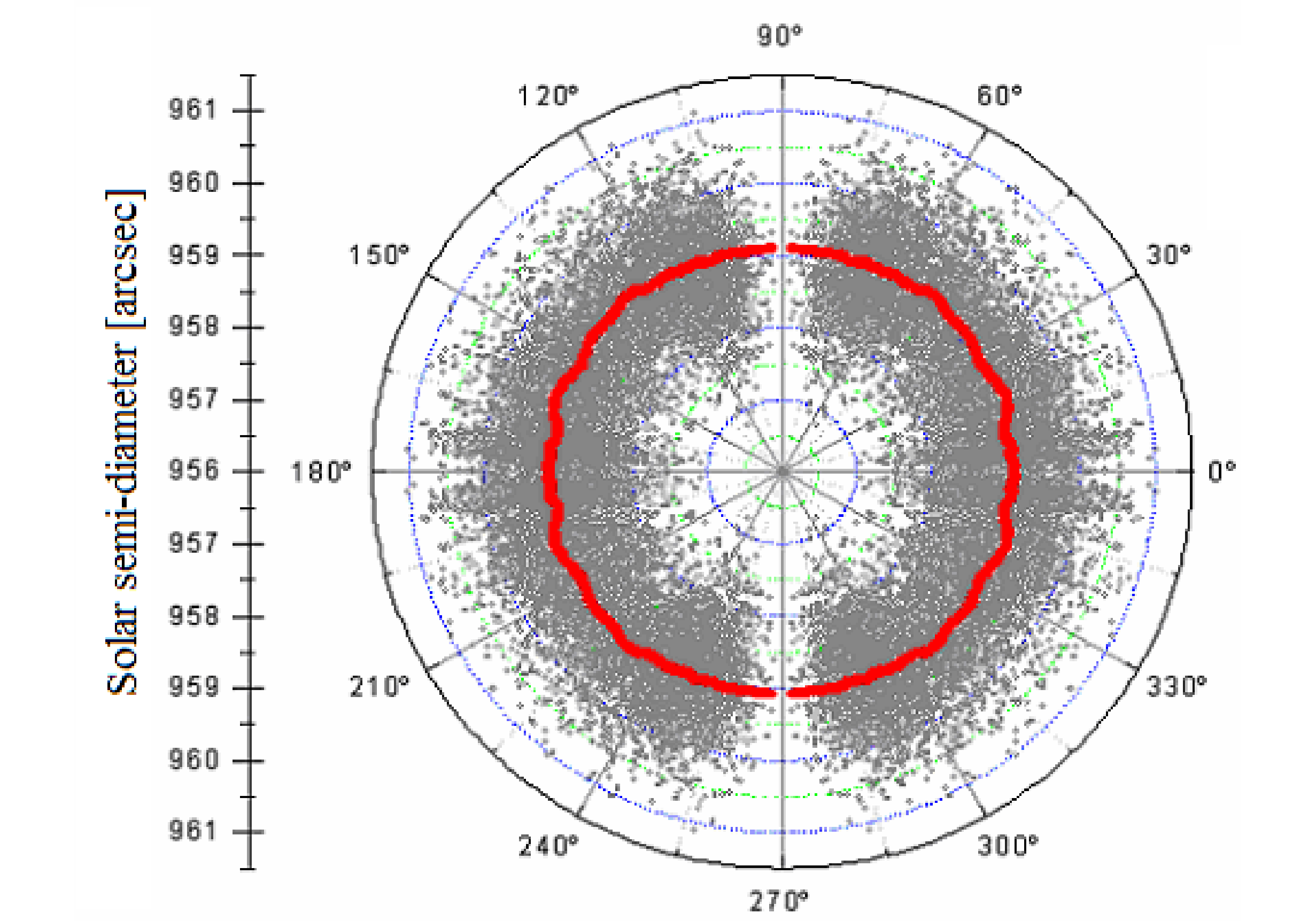}
\caption{The oblateness of the Sun 1998-2003 with Rio Astrolabe: average values in red.}
\label{fig:oblate}
\end{center}
\end{figure}

\begin{figure}[htb]
\begin{center}
\includegraphics[width=15cm]{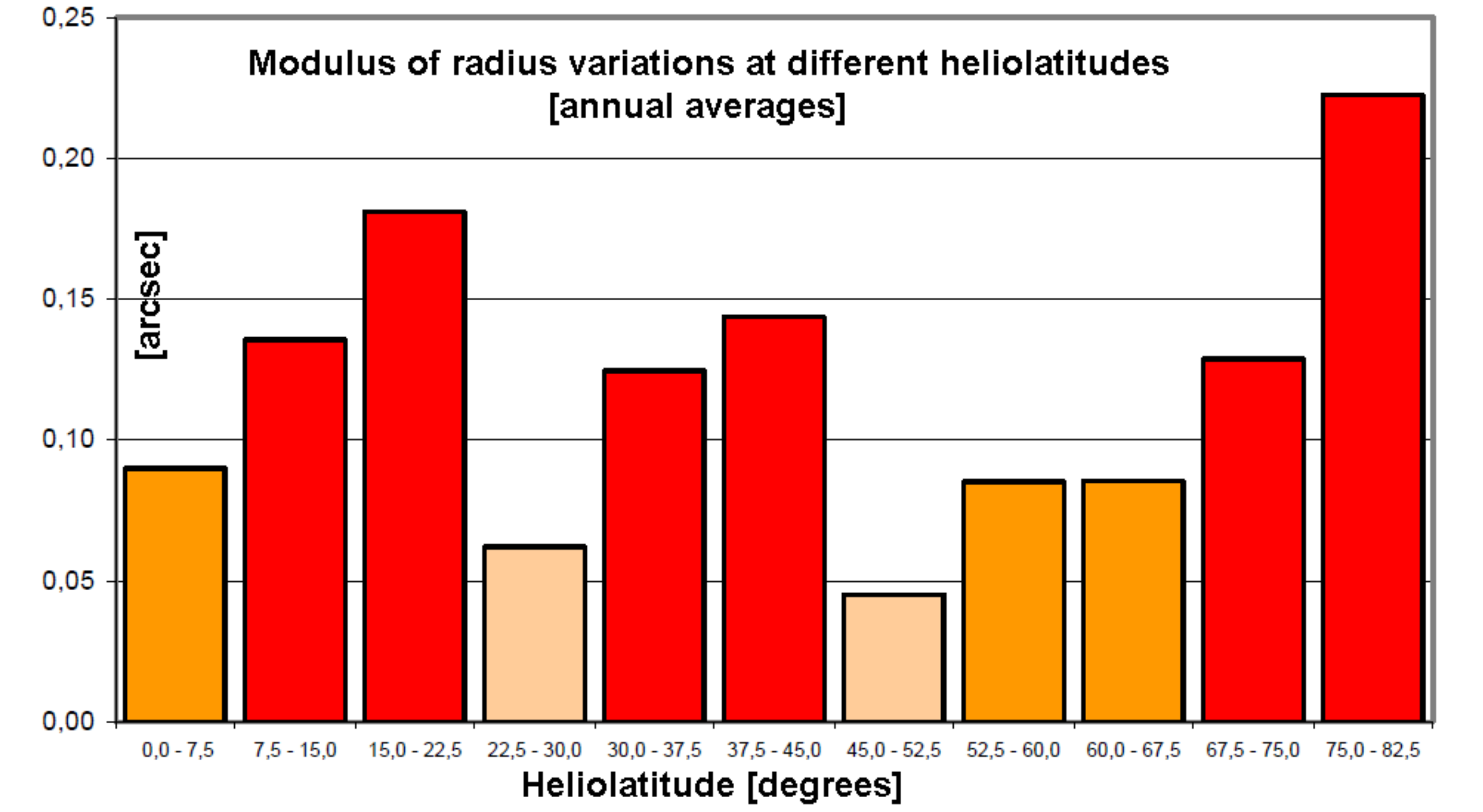}
\caption{The annual variations of averaged values as function of the heliolatitude [moduli].}
\label{fig:varia}
\end{center}
\end{figure}

\begin{figure}[htb]
\begin{center}
\includegraphics[width=15cm]{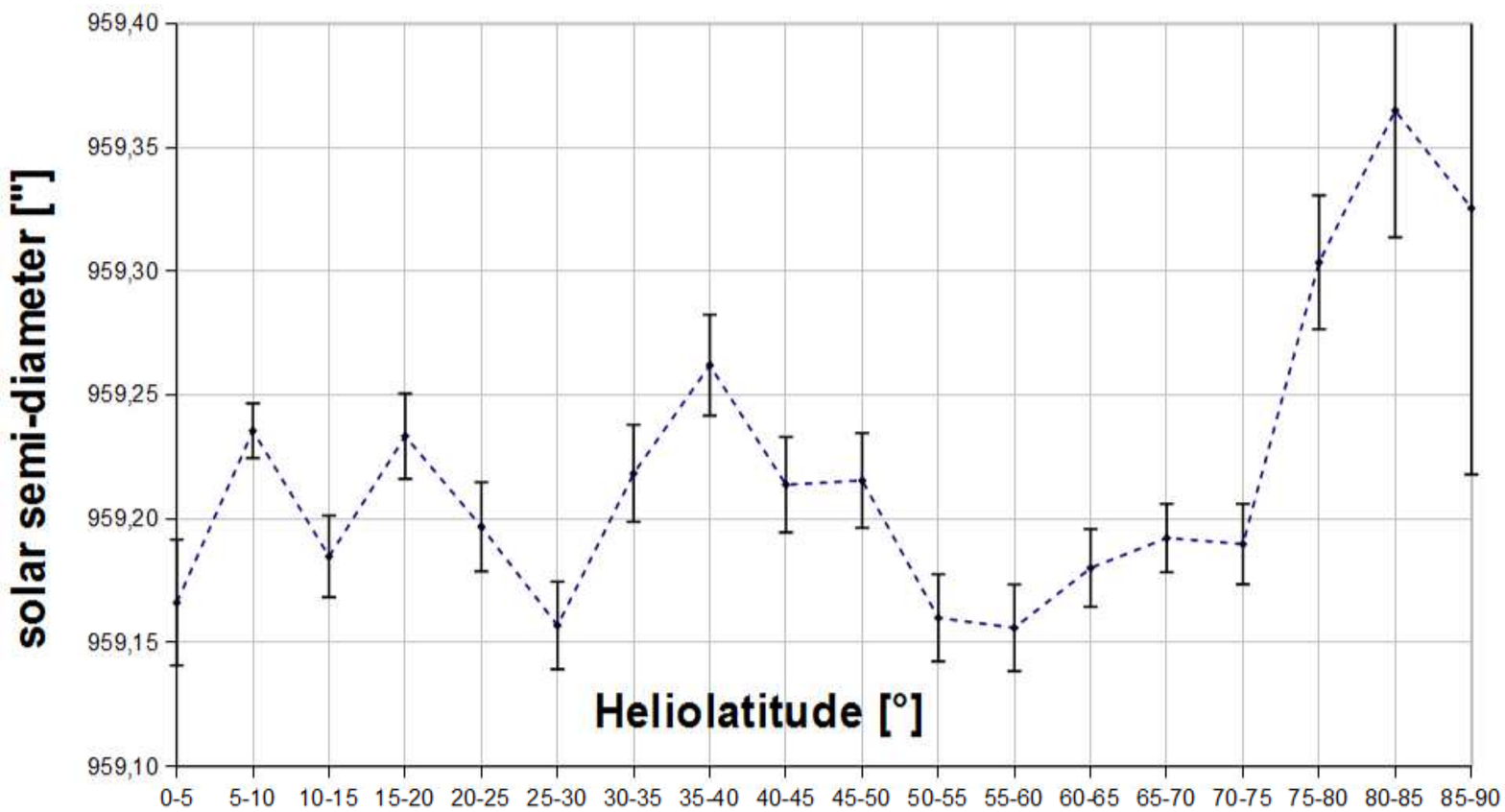}
\caption{The radius of the Sun at different heliolatitudes with statistical uncertainties.
The relative variations of radius are some parts in $10^{-4}$.}
\label{fig:due}
\end{center}
\end{figure}

\section{From the Astrolabe to the Reflecting Heliometer of Rio de Janeiro}

The Heliometer operating at the Observatorio Nacional in Rio de Janeiro has been conceived to obtain the measurement of the solar diameter at all heliolatitudes with an accuracy of one part over 100000. The first results (2010-2013) have been discussed in previous papers\cite{andrei13,sigismondi13} along with the data from the solar Astrolabe which observed from Rio during the period 1998-2009,\cite{sigismondi14} covering a whole solar cycle. 

The observations made with the Heliometer are characterized in the following ways:

\begin{itemize}
\item{the exposure time is around 1/1000 s, well below the typical time of
 the high frequency turbulence,\cite{berdja} which is freezed, being a short-pose exposure; the reduction of the diameter does not occur.\cite{berdja2}}
\item{the low frequency atmospheric motions does not have effect because the motion is freezed by the photo.}
\item{the vertical refraction acts heavily on the image and it is computed.}
\item{the horizontal refraction acts on the image and it is computed.}
\item{all heliolatitudes are measurable with the Heliometer through the rotation around its axis.}
\end{itemize}


The  Reflecting Heliometer of Rio de Janeiro is expected to improve rapidly the measurement of the oblateness when it will be fully operating.
These studies will cover also space and Earth climate issues, with the micro-variations of the solar diameter related to corresponding variations of the irradiance, and of coronal mass ejections.


\section{Conclusions}
The Sun is a self-gravitating body and the sphere is the best approximation of its figure.
The history of the measurements of the solar oblateness from ground is still in the phase of progressive approximations toward the real values.\cite{rozelot}
The evidence that the convergence is still ongoing arises from the comparison between the averaged values from one year to the other binned each $5^o$ of heliolatitude: in Fig~\ref{fig:sigma} the scatters are larger than the individual statistical uncertainties.
This is an evidence that something in the measurement process is still not fully understood.

Conversely the possibility to explain the differences in various measurements of the solar diameter with atmospheric sistematicities like UTLS\cite{desnoux} implies a sort of lensing effect which is constantly along the line of sight, difficult to be accepted.

The information about the variations of the asphericities in the Sun comes also from helioseismology.
These data and the observations of the variations of the solar diameter still does not match completely, even considering satellite data.
At solar minimum the asphericities $c_2$ and $c_4$ are expected to vanish, while at maximum activity the Sun surface is more corrugated, resulting a more efficient radiator. The corresponding relative variation of the solar diameter, according to helioseismology, should not exceed $10^{-6}$.\cite{dziembowski} Concerning direct measurements of the solar diameter and the oblateness, both satellite and ground-based data does not achieve yet this accuracy.
From the observational point of view the progress of the accuracy has been evidenced for the last techniques adopted in solar astrometry: both are present in the Observat\'orio Nacional in Rio de Janeiro and they are the Solar Astrolabe and the Reflecting Heliometer. 
Two millennia of atmospheric refraction models has been sketched to understand when the solar oblateness, hidden under the apparent flattening of the solar disk, particularly evident at the horizon, could have been measured.
The formulae of the differential refraction either vertical and horizontal were already used in 1786; the theory of atmospheric turbulence affecting images was developped in the last quarter of XX century,\cite{hill,berdja2} and the evidences of anomalous refractions and low frequency and large scale motions of the atmosphere was shown only in 2010.\cite{sigi-seeing} 

The expectations from the reflecting heliometer and other heliometric instruments ground-based is dependent also on the detailed model of the atmospheric refraction used in the data reduction. Space-based instruments are exempted from the refraction, but they suffer shorter lifetimes and larger thermal stresses which affect severely their measurements.
The link between the observations and the theories on the solar figure has also been discussed: a complete agreement is still lacking, because of unresolved problems in observations, and complicate stellar models at accuracy level of $10^{-6}$.
Space weather and Sun-Earth connections will be based on more precise data, 
when the standard of solar observation from ground will reach a few milliarcsec of accuracy, as it is expected
with the Reflecting Heliometer of Rio de Janeiro in duty.

\bigskip
Costantino Sigismondi acknowledges the CNPq PV grant 300682/2012-3.

\end{document}